\documentstyle[preprint,eqsecnum,aps,prb]{revtex}

\begin{document}
\draft
\preprint{  }
\title{ROUGHENING OF THE Cu(110) SURFACE}
\author{J. Merikoski, H. H\"akkinen\cite{byline}, M. Manninen and J. Timonen}
\address{Department of Physics, University of Jyv\"askyl\"a, P.O. Box 35\\
FIN-40351 Jyv\"askyl\"a, Finland}
\author{K. Kaski}
\address{Tampere University of Technology, P.O. Box 692\\
FIN-33101 Tampere, Finland\\and\\
Research Institute for Theoretical Physics, P.O. Box 9\\
FIN-00014 University of Helsinki, Finland}
\maketitle
\begin{abstract}
The structure of the Cu(110) surface is studied at high temperatures using
a combination of lattice-gas Monte Carlo and molecular dynamics methods with
identical many-atom interactions derived from the effective medium theory.
The anisotropic six-vertex model is used in the interpretation of the
lattice-gas results.
We find a clear roughening transition around $T_R = 1000$ K,
and $T_R/T_M = 0.81$.
Molecular dynamics reveals the clustering of surface defects as the
atomistic mechanism of the transition and allows us to estimate
characteristic time scales.
For the system of size 50x50, the time scale of the local roughening
at 1150 K of an initially smooth surface is of the order of 100 ps.
\end{abstract}

\section{Introduction}
\label{sec:intro}
Thermal disorder of fcc(110) metal surfaces has attracted considerable
theoretical and experimental interest during the past few years.
Due to its high structural and energetic anisotropy,
this surface  provides an intriguing case where a number of
competing disordering mechanisms can be found.\cite{Pa90,Ke91}
Roughly speaking, the (110) surfaces can be divided into two categories.
Metals such as Cu and Ni  preserve their bulk-terminated (1x1)
structure as the ground state,
consisting of nearest neighbour chains separated by
the conventional fcc lattice constant.
On the other hand, 5d metals Ir, Pt and Au have the tendency of forming
the so-called (1x2) missing-row reconstructed structure
or even further a (1xn) micro-faceted structure,
where only every n:th chain is left in the first crystal layer,
and the surface profile has (111) facets in the atomic scale.
For the non-reconstructed case, we can choose copper as the prototype.

Anomalous increase of disorder has been observed at Cu(110) between 500 K
and 1000 K by using X-ray\cite{Mo87} and helium diffraction\cite{Go85}
methods,
low-energy ion scattering and  inverse photoemission (IPE),\cite{Fa87}
low-energy electron diffraction (LEED),\cite{Th90}
high-resolution electron-energy-loss spectroscopy (HREELS)\cite{Ba91} and,
most recently,
the impact-collision ion-scattering spectroscopy (ICISS).\cite{Du91}
It is now generally accepted that strongly anharmonic vibrations of surface
atoms (anomalous Debye-Waller effect) are responsible for the increase
of disorder at the surface above 550 K.
It has been conjectured that these vibrations induce a roughening transition
where the lattice structure of the surface is preserved,
while the height-height correlations between the surface atoms diverge
logarithmically.
This was the initial conclusion from the X-ray diffraction experiments
by Mochrie,\cite{Mo87}
who estimated the lower bound of the transition temperature $T_R$ to be 870 K.
This result was later criticized by Zeppenfeld {\it et al.}\cite{Ze89} who
showed that the line shape of the specularly reflected thermal helium beam is
not consistent with a rough surface with logarithmic height-height correlations
up to 900 K.
By extending the measurements recently up to 1200 K, they find in fact that
$T_R=1070$ K.\cite{Ke92}
Also D\"urr {\it et al.}\cite{Du91} have recently observed structural
changes above 1000 K, the nature of which remains unclear.
It has been known for a long time that the (110) surface is missing from the
equilibrium crystal shape of copper just below
the melting temperature $T_M$,\cite{St76}
but the nature of the disordering mechanism has not been known for the
temperature range of 1000 K up to $T_M$.
Specifically, the possible existence of both the roughening and the surface
premelting transitions,\cite{Fr85} as well as their possible mutual
interactions, have not been resolved.

The fcc(110) face is the most open one among the low-index faces of fcc
lattice, with a structure resembling the (more open) stepped (113),(115),...
surfaces.
The basic excitation of a stepped surface is a kink on the step edge,
whereas at a low-index face,
roughening is induced by formation of vacancies and adatoms or
excitations with a still higher energy.\cite{Be87}
In the limit of very large anisotropy,
the anisotropic BCSOS (body-centered solid-on-solid) model of
the fcc(110) face and the terrace-step-kink model
of high-index surfaces become equivalent.\cite{Sa88}
In the case of copper (110) surface, anisotropy is not very large,
and its roughening is best described by that of a low-index face.
This kind of roughening has been described by various Monte Carlo
simulations on lattice-gas and solid-on-solid models.\cite{Be87}
Recently, molecular dynamics simulations using many-atom interactions have
shown pronounced disorder and finally premelting of the (110) surface
of Al, Ni, Au and Cu.\cite{St88,Ha92}
Identification of the rough phase is difficult in the molecular dynamics
simulations, however, because tractable sample sizes and simulation times
preclude the identification of a logarithmic behaviour of the
height-height correlation function.

In this paper we use a new theoretical approach to investigate the
roughening of the Cu(110) surface.
By extracting the interatomic interactions from the well-tested
effective-medium theory,\cite{Ja87}
the structure of the Cu(110) surface at high temperatures is studied by
means of both the lattice-gas Monte Carlo (LGMC) and molecular
dynamics (MD) simulations with comparable system sizes.\cite{He86}
A brief account of the results was already given in Ref.~\onlinecite{Ha93},
but here we shall give a detailed description of the method and of the
various results it can provide.

In the LGMC simulations we identify the rough phase from the behaviour of
the height-height correlation function
\begin{equation}
G({{\vec r}}) =
   \langle ( h({{\vec 0}})\!-\!h({{\vec r}}) )^2\rangle\ ,
\label{eq1}
\end{equation}
where $h({{\vec r}})$ is the height of the surface at
point ${{\vec r}}$=$(x,y)$.
Below the roughening temperature $T_R$, the interface width is finite with a
finite correlation length,
while at and above $T_R$ the correlation function diverges logarithmically,
\begin{equation}
G({{\vec r}}) \sim 2A(T)\ln r\ +c\ ;\ \ \ \ T\ge T_R\ .
\label{eq2}
\end{equation}
At the transition point the coefficient $A$ has the universal
value $A(T_R)$=1/${\pi}^2$.\cite{We80}
Using the LGMC model we find a clear roughening transition around 1000 K.
The dynamics of the rough surface is studied via MD simulations employing
the same interactions as used in the LGMC method.
The LGMC results are qualitatively similar to the MD results which include the
effects of lattice vibrations and relaxation.
Furthermore, the atomistic mechanism of the roughening is found through MD
to be the creation of defect clusters of vacancy and adatom type,
and the relevant time scales can also be estimated.
It is noteworthy that the estimated $T_R$ is, for the first time,
in a realistic temperature range in comparison with the previously
calculated (by MD) surface premelting and bulk melting temperatures for
the same interatomic potential.\cite{Ha92}

The paper is organized as follows.
Section \ref{sec:model} describes the models which have been employed,
and gives a technical description of our simulations,
including also a discussion of the effects of the inherent differences between
our LGMC and MD models.
The determination of the roughening temperature by LGMC simulations
is described in Section \ref{sec:resmc}.
In Section \ref{sec:resmd} we discuss the results of the MD simulations,
found by following the evolution of the surface from a smooth to a
locally rough phase.
We follow in particular the time evolution of the occupation numbers of the
surface layers and of the cluster distributions, and atomic diffusion.
Our conclusions are given in Section \ref{sec:concl}.

\section{Models and methods}
\label{sec:model}
The effective medium theory (EMT) is an approximative method for calculating
the total energy of an arbitrary arrangement of metal atoms.\cite{Ja87}
EMT relates the energy of an atom in metal to the local electron density and,
therefore, implicitely includes many-atom effects.
It gives a reasonable description of a number of surface properties
including surface relaxation, reconstruction and phonon
spectra.\cite{Ja87,Ha91}
We have extended the interaction range in EMT to the third nearest neighbours
in the fcc copper lattice as outlined in Ref.~\onlinecite{Ha92}.
These interactions have been used in both the lattice-gas Monte Carlo and
the molecular dynamics simulations.

\subsection{Lattice-gas Monte Carlo simulations}
\label{sec:metmc}
In our Monte Carlo simulations we use a face-centered cubic lattice-gas model
with many body interactions derived from EMT in the form of a function
$E$=$E(C_{1},C_{2},C_{3})$,
where $C_j$ is the number of the $j^{th}$ neighbours for an atom in the
lattice.
In Fig.~\ref{fig1} we show $E$ as a function of $C_1$ and $C_2$
with $C_3=2C_1$.
The broken curve is the derivative $dE/dC_1$ or the energy cost of a
broken nearest neighbour bond.
We also show in this figure
the simplest possible pair interaction model (dotted line)
that is normalized such as to produce the right cohesive energy in a perfect
lattice and zero energy at zero coordination.
The basic excitation energies at Cu(110) are about 2.5 times higher in the
pair interaction model than in the EMT.

The EMT lattice-gas Hamiltonian can be written in the form
\begin{equation}
H-\mu N = \sum_i n_i\lbrack E(C_{1i},C_{2i},C_{3i})
            -v_i -\mu\rbrack\ ,
\label{eq3}
\end{equation}
where $n_i=0,1$ is the occupation number of the lattice site $i$,
$C_{ji}$ is the number of its $j^{th}$ neighbours
and $\mu$ is the chemical potential.
The external potential $v$ is needed for the formation of an interface in the
lattice-gas model.\cite{Ol78}
We have assumed the form $v_i=a$ for the first layer, and $v_i=bz_i^{-3}$
for the other layers, where $z_i$ is the distance of the lattice site $i$
from the 'substrate' or, equivalently, the layer number ($z_i=1,2,..$).
With our choice $a=6000$ K and $b=9000$ K the formation of the first
'adsorption layer' is essentially two-dimensional and
the bulk limit is reached for $z_i\simeq $10.

For comparison with our lattice gas results, we have also mapped the lattice
gas interface to an anisotropic six-vertex solid-on-solid model
or the BCSOS model.\cite{Be77,Ja84}
A schematic picture of the fcc(110) face seen from above,
and its mapping to the six-vertex model, are shown in Fig.~\ref{fig2}.
In Fig.~\ref{fig2}a, the black dots denote surface atoms at level 0 and the
open symbols with $+$ and $-$ are surface atoms at levels $+1$ and $-1$,
respectively.
The corresponding vertex configuration is also shown.
The energy parameters of this model are determined from the surface excitation
energies of EMT.
The energy cost of broken bonds,
$\epsilon_1=\epsilon_2$ in the nearest neighbour direction
and $\epsilon_3=\epsilon_4$ in the next nearest neighbour direction,
are found from the energy function $E=E(C_1,C_2,C_3)$ for a coordination
corresponding to the uppermost atom layer
for which $C_1\simeq$ 7 and $C_2\simeq$ 4.
These energies are then scaled such as to reproduce the right energy for an
adatom-vacancy pair on an otherwise flat face.
In this way we find for the vertex energies: $\epsilon_1=\epsilon_2=$ 1485 K
(the derivative of $E$ with respect to $C_1$),
$\epsilon_3=\epsilon_4=$ 322.5 K
(the distance between the curves for different $C_2$ in Fig.~\ref{fig1}),
and $\epsilon_5=\epsilon_6=0$.
The model is in fact a pair potential model with an energy scale derived from
a many-atom model.
Notice that no external potential is needed here to define the location of the
interface.

The exact transition temperature $T_R$ in the thermodynamical limit
of the six-vertex model is given by\cite{Ba89}
\begin{equation}
\Delta(T_R)=-1;\ \ \ \Delta(T)=(a^2+b^2-c^2)/2ab\ ,
\label{eq4}
\end{equation}
where $a=\exp(-\epsilon_1/kT)$, $b=\exp(-\epsilon_3/kT)$ and
$c=\exp(-\epsilon_5/kT)$.
Using the energy parameters given above, the roughening temperature of the
EMT six-vertex model in the thermodynamical limit is
$T_R^{6v}(L\!\!\to\!\!\infty)=$ 1090 K.
The behaviour of the coefficient $A$ defined in Eq.\ (\ref{eq2}) is also known
exactly\cite{Yo80} as a function of temperature,
\begin{eqnarray}
  A(T) & = & \lbrack \hbox{$1\over2$}\pi^2-\pi\arcsin\Delta(T)\rbrack^{-1}
\label{eq5} \\
       & = & 1/ \pi^2 +(T-T_R)^{1/2}\lbrack a_1+a_2(T-T_R)+...\rbrack   \ .
\nonumber
\end{eqnarray}
In Eq.\ (\ref{eq5}), the value of $A$ at the transition point,
$A(T_R)=1/ \pi^2$, is universal.
We will use the leading square root (Kosterlitz-Thouless) behaviour near
$T_R$ of coefficient $A$ in interpreting the results of our
many-body-lattice-gas Monte Carlo calculations.

It is evident from Eq.\ (\ref{eq4}) that the critical temperature of the
six-vertex model depends on two parameters,
the energy scale parameter $\epsilon=\epsilon_1 - \epsilon_5$,
and the anisotropy of vertex energies
$\nu=(\epsilon_3 - \epsilon_5)/(\epsilon_1 - \epsilon_5)$.
In our parametrization the anisotropy ratio is $\nu=$ 0.217.
In Ref.~\onlinecite{Tr89} the vertex energies were determined from surface
energies $\sigma_{110},\sigma_{100},\sigma_{111}$,
and the anisotropy ratio was found to be
$\nu=(\sigma_{111}\sqrt{3/2}-\sigma_{110})/
(\sigma_{100}\sqrt{2}-\sigma_{110})$.
In our EMT lattice-gas model this method of defining the vertex energies
would give $\nu=$ 0.141.
This value is still much higher than the value $\nu=$ 0.080
obtained in Ref.~\onlinecite{Tr89} from the
relaxed surface energies of the embedded atom method.
In that work the low $\nu$ value leads to a much lower estimate for the
critical temperature.
This discrepancy is mainly caused by the difficulty in
including in the energy scales of a pair interaction model
the many-body effects of the true interactions.
Simulations with realistic many-atom interactions are thus really needed to
settle this question.
We shall show below that,
just above $T_R$, typical excitations at the Cu(110) surface are not
large (100) or (111) facets,
but monoatomic steps with a temperature dependent kink density.
This behaviour indicates that the effects
on the surface excitation energies
of both many-atom forces and relaxation are very complicated.

We use the method of equilibrium Monte Carlo simulations,
which is known\cite{Mu79} to be suitable for lattice-gas and solid-on-solid
models.
The many-body nature of the Hamiltonian Eq.\ (\ref{eq3}),
and the extension of the
interactions up to third neighbours in the lattice-gas model, increases the
computation time by approximately a factor of 20 in comparison with the pair
potential model we have also used.
The lengths of the runs varied from a few thousand (for isotherms)
to 500000 Monte Carlo steps per lattice site,
for systems of size L$^2$ with 12$\le$L$\le$196.

\subsection{Molecular dynamics simulations}
\label{sec:metmd}
The lattice-gas model discussed above lacks two major physical features.
First, there is no real physical time connected with the evolution in the
lattice.
Second, the effects of lattice vibrations and surface relaxation are ignored.
To study these effects we have performed large-scale molecular dynamics
simulations employing the same interaction potential
from which the site-site interaction energies are calculated for the LGMC
model.
This potential is fully documented in a previous work on surface
premelting.\cite{Ha92}

Our sample consists of 8 dynamical (110) layers on top of 4 static substrate
layers,
with either 30 close-packed $\lbrack  1\bar 10\rbrack$ rows of 30 atoms or 50
rows of 50 atoms.
The number of dynamical atoms is then 7200 and 20000 in the smaller and
the larger sample, respectively.
The dynamics is realized by the Nos\'e-Hoover canonical thermostat.\cite{No85}
This method is a natural choice in our simulations since,
throughout this work, we wish to compare the molecular dynamics results
with those of the LGMC model,
the lattice configurations of which are generated at constant temperature.
The lattice parameter of the substrate is adjusted such as to correspond to the
simulated temperature as defined from the thermal expansion produced by
the interaction potential.
The equations of motion are integrated by using the velocity-Verlet
algorithm,\cite{Sw82}
which has been modified to handle the thermostat equations.\cite{Ha92}
Due to the good stability of the algorithm, we have been able to use a
large time step of 14 fs,
which is roughly a tenth of the Debye period for copper.
The large number of atoms in MD requires special methods for
selecting the pairs of atoms for which the interactions are calculated.
The method we have used is described in the Appendix.

The results which we shall report in this paper are those for temperatures
$T$ = 1100 K and $T$ = 1150 K.
In the $T$ = 1150 K case we shall analyse a pair of runs,
one initiated from a rough lattice-gas configuration generated by
the LGMC model with a number of atoms corresponding to full layers
(run A, 30x30 sample, length 406 ps),
and the other initiated from an  undefected smooth surface
(run B, 50x50 sample, length 616 ps).
At $T$ = 1100 K we have made only a single run (run C, 50x50 sample, length
560 ps), initiated from the undefected smooth surface.
The reason for studying these temperatures is that,
according to the LGMC results,
they are within the range where the surface is rough but not yet premelted
(the surface premelting of Cu(110) occurs in our MD model
at $T\ge 1200$ K\cite{Ha92}).
We have also made some simulations at lower temperatures,
but due to the slowing down of diffusion,
the statistics for the defect clusters becomes poor.

\subsection{Inherent differences of the LGMC and MD models}
\label{sec:inher}
In our LGMC simulations the energy function $E$ is derived from the
effective medium theory by using the bulk lattice constant at zero
temperature.
In the lattice-gas model there is no surface relaxation,
but its effect on the transition temperature can easily be estimated
by comparing the energy of the basic excitation of the unrelaxed
lattice-gas model with that at the relaxed surface.
The inward relaxation\cite{Ha92} means higher electron density
or effectively higher coordination,
and thus decrease in the slope of the energy function of Fig.~\ref{fig1}.
In our lattice-gas model we therefore {\it overestimate} the excitation
energies and,
consequently, the roughening temperature of Cu(110), by about 5\%.
This estimate will be used in determining our final estimate for $T_R$.

In the LGMC energy function the effect of thermal expansion is not included,
but this effect is in most part included in our estimate of the relaxation
effects on the roughening temperature given above.
A more fundamental difference between the two models is the fluctuating
particle number in the LGMC simulations vs.\ the fixed particle number
in the MD simulations.
In practice this difference leads to a much slower sampling of uncorrelated
lattice configurations in the MD simulations.
In the MD simulations the interaction potential has been smoothened at the
cutoff radius (between the $3^{rd}$ and $4^{th}$ neighbours) as described
in the Appendix of Ref.~\onlinecite{Ha92}.
This has been done in such a way that effectively (through indirect
interactions) increases the anisotropy, and thus lowers the roughening
temperature of our MD model.
We only note here that the way of cutting the interaction range in MD
simulations can have a considerable effect on such surface properties
that are very sensitive to the strength of the interaction beyond the
nearest neighbours.
This is evident from the discussion of the missing-row reconstruction given
in Ref.~\onlinecite{Ha91}.

\section{MC results}
\label{sec:resmc}
Calculated height-height correlation functions for the lattice-gas system with
L = 96 and for the six-vertex system with L = 50 are shown in Fig.~\ref{fig3}.
Indicated symbols denote the Monte Carlo results and solid lines are best
logarithmic fits to the data points.
The distance $r$ is in the direction of the next-nearest neighbours in the
surface plane.
In the lattice-gas model the value of the chemical potential was chosen such
as to produce an interface at a level $z_i> $ 10, i.e.\ $v_i < $ 10 K.
For temperatures above $T$ = 1000 K,
the effect on the surface structure of the substrate potential
difference between adjacent layers is negligible.
As noted in Ref.~\onlinecite{Sh78},
finite size effects become significant for $r\sim$ L/5.
On the other hand, a few lowest values of $r$ have been discarded in the fits
because Eq.\ (\ref{eq2}) becomes exact only at large $r$.
We have also done simulations for bigger systems,
but increasing computational cost makes it impossible to sample the
long-wavelength fluctuations properly.
This finite time effect prevents us from being able to make fits to the
correlation functions at larger values of $r$.

{}From the logarithmic fits to the correlation functions we can infer the value
of the coefficient $A$ defined in Eq.\ (\ref{eq2}).
The result is shown in Figs. \ref{fig4}a and \ref{fig4}b as a function of
temperature for the
EMT lattice-gas model and for the six-vertex model, respectively.
A square root fit of Eq.\ (\ref{eq5}) to the Monte Carlo data is shown by the
solid line.
{}From these fits we find $T_R^{lg}$ = 1000 K and $T_R^{6v}$ = 1030 K.
It is evident that the dependence on the system size
of the transition temperature is beyond our accuracy.
The difference between the exact value in the thermodynamic limit of
the six-vertex model ($T_R^{6v}({\rm L}\!\!\to\!\!\infty)$ = 1090 K)
and our numerical result ($T_R^{6v}$ = 1030 K) is caused by two effects,
the effect of finite size and (mainly) the relatively short distances
over which the fitting of the correlation function has been made.
A systematic deviation of similar magnitude ($\sim$ 60 K) is expected
to be present also in our lattice-gas result,
where MC runs of comparable length have been performed, and logarithmic
fits to the correlation functions have been made in a similar way.
Thus we can conclude that $T_R^{lg}({\rm L}\!\!\to\!\!\infty) \approx$ 1060 K.

For comparison we have also determined the $T_R$ of our lattice-gas model
by studying the finite size behaviour of the interfacial width ,
$\delta h^2 = \langle ( {\bar h}\!-\!h({{\vec r}}) )^2\rangle_{{{\vec r}},t}$
\ \ ,
where ${\bar h}$ is the nominal surface height for a given
configuration.\cite{Ma92,Sw77}
Above the transition temperature, $\delta h^2$ diverges logarithmically as a
function of the linear size of the system size, L,
or $\delta h^2\sim A(T)\ln$L for large L.
In Fig.~\ref{fig4}c we show the result for the transition temperature
as determined from the coefficient $A(T)$ for the interfacial width.
Here data for L = 12, 18, 24 and 30 has been used.
The result is found to be consistent with that obtained from the correlation
functions.

The Hamiltonian Eq.\ (\ref{eq3}),
leads us to still another way of deducing the roughening transition,
namely through layering transitions in an adsorption system.\cite{Bu51}
For a thickening adsorption film the critical points of the layering
transitions approach the bulk roughening temperature.\cite{Ol78}
In Fig.~\ref{fig5} we show a few adsorption isotherms corresponding to the
formation of the first 'adsorption layer' and
the inverse slope of the adsorption isotherms at the layering transition
point, $\mu=\mu_1$, as a function of temperature.
The solid line in Fig.~\ref{fig5}b is a fit with the two-dimensional
critical exponent of the susceptibility, $\gamma = 7/4$,
and with a critical temperature $T_1 = 845$ K.
A similar analysis for the next layer gives $T_2 = 870$ K.
We do not intend to determine $T_R$ by repeating this procedure for further
layers;
we only note here that these critical layering points, $T_k$, are lower bounds
for the roughening temperature.\cite{Ni84}
In our lattice-gas model the first adlayer grows essentially two-dimensionally
(the second layer is almost unoccupied), which means that
$T_1\approx T_{2d}^{lg}$ and we have $T_{2d}^{lg} / T_R^{lg}$ = 0.80.
By using vertex energies $\epsilon_1$ and $\epsilon_3$ as the energy parameters
in the two-dimensional Ising model,\cite{Ba89}
we obtain $T_{C}^{Ising} / T_R^{6v}$ = 0.80.
Besides being a further confirmation of our estimate of the roughening
temperature in the lattice-gas model, this result shows
that a simple pair-potential model with relatively short-range interactions
(such as the six-vertex model) can be used to describe the roughening
transition of a metallic system,
provided the values of the energy parameters of the model are determined in an
appropriate way.
The effect on the roughening transition
of the broken particle-hole (adatom-vacancy) symmetry\cite{Ry86,Ni90}
is too small to be detected in our lattice-gas model:
the energy difference between convex and concave corners on Cu(110) is
less than 10 K.

In Fig.~\ref{fig6} snapshots of typical configurations of a lattice-gas
system with L = 30 are shown for two temperatures.
Atoms in the lowermost odd layers are coloured black to guide the eye.
Below $T$ = 1000 K excitations are mainly single adatoms and vacancies
at the surface.
Around $T=T_R$ the connectivity of the adatom and vacancy clusters increases,
and at higher temperatures the surface is clearly 'rough'.
The anisotropy of the clusters reflects the high anisotropy of the excitation
energies in the surface plane,
but there is no clear indication of (111)-type facetting
since the anisotropy in our model is not very large.
In the snapshot at $T=1100$ K two overhangs can be seen:
in both cases one of the four nearest neighbours at the next atom layer
below is missing.
Near $T=T_R$ the overhang density is very low ($<$0.001),
and the solid-on-solid approximation of the six-vertex model is valid.

We conclude that the roughening temperature of our EMT lattice-gas model
is $T_R^{lg}$ = 1060 K.
Considering the effect on the energy scale of surface excitations of
relaxation (see Sec.\ \ref{sec:inher}), our best estimate for the roughening
temperature of Cu(110) is $T_R\approx$ 1000 K.
Comparing this with the bulk melting point for the same potential,
$T_M$ = 1240 K,\cite{Ha92} we find $T_R/T_M$ = 0.81.

\section{MD results}
\label{sec:resmd}
We shall first compare the runs A (rough initial configuration) and B
(smooth initial configuration) by studying the occupation
numbers of the surface layers as a function of time.
Atoms are attached to layers according to their z-coordinate by dividing
the z-dimension of the sample in slices of thickness
$\Delta z= a/2\sqrt{2}$,
where $a$  is the conventional fcc cubic lattice constant.
$\Delta z$ then corresponds to the distance between adjacent (110) planes in
the bulk.
We adopt hereafter the convention of 'crystal' and 'adatom' layers.
The crystal layer $c_1$ is the uppermost full layer of the undefected surface
at $T=0$, and the  adatom layer $a_1$ is the first unoccupied layer at $T=0$.
The other crystal and adatom layers are numbered from the $T=0$ interface
into the bulk and into the vacuum, respectively.

The occupation numbers of the surface layers in runs A and B for $T$ = 1150 K
are plotted in Fig.~\ref{fig7} as a function of time.
We can extract from Fig.~\ref{fig7} several important features of surface
roughness.
First, the initial roughness (i.e.\ occupation numbers) of the LGMC surface in
the run A remains essentially the same also during the MD phase.
The width of the interface seems to slightly increase after the MD
simulation is started,
which can be seen from the systematic increase in the occupation of the two
adatom layers.
The rough initial LGMC configuration is thus stable against the lattice
vibrations and relaxation effects,
i.e., it has  no  tendency of smoothing during the MD simulations.
Second,  we can reach the  roughness of the run A during the first 150 ps
of run B, which is initiated from the smooth surface.
Notice that this is directly an estimate for the physical time needed for
the surface to 'roughen' locally via its intrinsic dynamics.
Third, one can see irregular fluctuations of a few tens of ps,
on top of the normal statistical fluctuations.
These long-wavelength fluctuations can especially be seen in the run A,
which is made for the smaller, 30x30, sample.
These fluctuations are connected with the diffusion dynamics of adatom and
vacancy clusters.
They show that, during our simulation time, it is possible to
produce at least a few uncorrelated lattice configurations, i.e.,
we are able to travel through the configuration space of the rough surface.

At $T$ = 1100 K (run C) the behaviour of the occupation numbers is
qualitatively similar to that shown in Fig.~\ref{fig7}.
However, it takes much more time (about 300 ps) to reach the average LGMC
level.
We can interpret this as  a 'critical slowing down' effect,
which makes the observation times for the cluster dynamics in MD simulations
extremely long when the transition temperature is approached.
Some snapshots of two-dimensional adatom clusters in layer
$a_1$ at 1150 K (run B) are shown in Fig.~\ref{fig8}.
The configurations in the first two snapshots (with a time interval of 14 ps
or 1000 MD time steps) are clearly strongly correlated.
In the third snapshot we obviously see a new lattice configuration.
Despite the large vibrations and atomic diffusion,
the clusters have a well-defined (110) surface symmetry.

In order to study atomic mobility we have calculated the probability
distributions $p(x,y,z;t)$ for the surface layers, where
$p$ is the probability that an atom is in the  position $(x,y,z)$
at time $t$, when it is at the origin at $t=0$.
The function $p(z;t)$ for layer $a_1$ shown in Fig.~\ref{fig9} illustrates
the large number of layer changes during the simulations:
After the observation time of 70 ps it is about equally probable to find the
atom in layer $c_1$ as in layer $a_1$.
This result is be related to the initial rising time of the occupation of the
layer $a_1$ shown in Fig.~\ref{fig7}.
The $p(x;t)$ distribution for the cross-channel $\lbrack 001\rbrack$ direction
is shown in Fig.~\ref{fig10}.
The spacing between the maxima in $p(x;t)$ corresponds to a half of the lattice
constant,
which indicates that the cross-channel $\lbrack 001\rbrack$ diffusion
is the main mechanism of layer changes.
This result has also been found from previous MD simulations of the fcc(110)
surface diffusion.\cite{Mr81}

We have studied the residence time of an atom at the lattice sites of each
surface layer by analyzing the atomic displacements during
the last 42 ps of the run B ($T$ = 1150 K), and by collecting the statistics of
jumps from one lattice site to another.
The collected distributions of the residence times
for the surface layers are shown in a
semilogarithmic scale in Fig.~\ref{fig11}.
After an initial rise all distributions seem to obey an exponential law,
$p\propto e^{-t/ \tau}$,
from which the mean residence times, $\tau$, for the diffusing atoms can be
estimated to be 1.5 ps , 2.2 ps, 3.5, ps and 4.5 ps  for $a_2$, $a_1$, $c_1$,
and $c_2$, respectively.
These residence times  correspond to 10-30 lattice vibrations.
Notice, however, that in layer $c_2$ 27\% of the atoms are not
diffusing at all during our observation time.
The 'true' residence time for that layer must therefore be longer
than the 4.5 ps quoted above.
The corresponding fraction of immobile atoms in $c_1$ is only 2\%,
and practically all atoms in layers $a_2$ and $a_1$ have made at least one jump
during our observation  time.

The diffusion constants $D_x$,$D_y$ for the surface layers are determined
from the time-dependent  mean-squared atomic displacements (msd)
$r^2(t)=\langle ({\bf r}(t)-{\bf r}(t_0))^2 \rangle$,
where the brackets mean averages taken over 20 initial times $t_0$ and
over atoms belonging initially to a given layer.
$D_\nu$, $\nu=x,y$,
can be obtained from  the Einstein relation as, for $t\rightarrow\infty$,
$r_{\nu}=2D_{\nu}t$+const.
We find that
($D_x,D_y$) for $a_2,...,c_2$ are (1.80, 4.13), (2.05, 2.69),
(1.71, 2.77), and (0.88, 1.24) $\times 10^{-5}$ cm$^2$/s, respectively.
It is evident that the calculated surface mobility is high,
near the mobility level in the bulk liquid at the melting point.\cite{Ha92}
However, the persistence of the lattice structure makes the diffusion
clearly anisotropic.
We emphasize that what has been considered here is the {\it intrinsic}
(tracer) diffusion constant, not the mass transport constant
which is expected to  become isotropic at $T$ = $T_R$ due to the macroscopic
step diffusion.\cite{Pa90,Ke91,Go83}

The time scale of the roughening of a smooth surface can also be measured
by studying the time evolution of cluster size distributions.
In Fig.~\ref{fig12}a we show the size distribution at $T$ = 1100 K
of adatom clusters in layer $a_1$ as a function of time (MD run C).
After a simulation of 0.56 ns we are still far from the equilibrium
distribution for clusters of size $N >$ 60.
At $T$ = 1150 K the size distribution is stable (within our statistical
accuracy) after 0.16 ns for $N$ up to $N \simeq$ 120.
The result for the time interval 160 ps to 616 ps (run B) is shown
in Fig.~\ref{fig12}b.
The dotted line is the behaviour of the lattice-gas system with of same size
and at the same temperature.
The results for the vacancy clusters in layer $c_1$ are similar.
The height-height correlation function $G$ (not shown here) of
Eq.\ (\ref{eq1}) as determined
from the MD data for $T$ = 1150 K, is an increasing function of $r$
only for $r \le 5$.
With $r$ taken to be in the direction of the next-nearest neighbours (the
direction of weaker bonds), this result reflects the fact that,
during this simulation of length 0.6 ns and well above the roughening
temperature, we have been able to collect reasonable statistics for defect
clusters only for $N \le 100$.
The dynamics of the cluster size distributions in our MD simulations is
consistent with the results for the kinetics of the roughening of a stepped
surface discussed in Ref.~\onlinecite{Se87}:
roughness sets in over short distances and then spreads asymptotically
as $t^{1/3}$.
The time scale of roughening,
100 ps (Fig.~\ref{fig7}) at $T$ = 1150 K and
300 ps at $T$ = 1100 K,
defined here as the time required for the occupation numbers of the innermost
layers to saturate to their 'rough' values,
clearly describes the {\it local} behaviour.
For systems of size $L^2$ with $L \ge 50$, this quantity should depend only
weakly on $L$.
We have not tried to determine the asymptotic behaviour for late times.

Based on the calculated diffusion constants in the surface region,
we can conclude that
for an atom found initially in layer $c_1$,
it takes at $T=1150$ K about 30 ns (50 $\times$ our simulation time)
to travel through a distance similar to the length of the simulation box in
the shorter y-direction.
The corresponding time for diffusion across the box in the x-direction
is about 90 ns.
These characteristic times provide an order-of-magnitude estimate for the
amount of computation time needed
for a reliable calculation of the height-height correlation
function at one temperature by the MD method.
In order to be able to determine $T_R$, a few temperatures below 1150 K
should be studied.
The computation time required by the present MD model, with the EMT
interactions for Cu(110), would be of the order of one year of Cray X-MP
CPU time.

\section{Conclusions}
\label{sec:concl}
We have studied the roughening of the Cu(110) surface via atomistic
simulations using both the Monte Carlo and molecular dynamics methods
which employ  the effective-medium theory as the interaction potential.
The Monte Carlo simulations made for our lattice-gas model,
adjusted by corrections arising from relaxation and finite size effects,
show clearly that the Cu(110) surface has a roughening transition
around $T$ = 1000 K,
about 200 K below the surface premelting temperature,\cite{Ha92}
determined for the same potential.
The rough phase has been identified from the logarithmic behaviour of the
height-height correlation function and of the interface width.
The transition temperature is obtained from the
temperature-dependent coefficient of the logarithmic term.
The finite size effects have been estimated by mapping the lattice-gas
interface to an anisotropic six-vertex model,
for which the behaviour in the thermodynamical limit is known exactly.
Knowing the bulk melting point for the same potential,
we find $T_R/T_M$ = 0.81.
This is in excellent agreement with the experimental result
$T_R/T_M$ = 0.79 given in Ref.~\onlinecite{Ke91}.
Molecular dynamics simulations of the same surface cannot produce
statistics enough for the height-height correlation function,
but they show the stability of the LGMC surface against lattice vibrations
and relaxation, i.e.,
the rough LGMC surface shows no tendency of smoothing when used as an initial
configuration for MD simulations.
Furthermore, MD simulations show that the roughening mechanism is
connected with the dynamics of diffusive adatom and vacancy clusters and
gives information about the corresponding time scales.
For the system of size 50x50 studied in this work,
an initially smooth surface 'roughens' locally in the time scale of
about 100 ps at $T$ = 1150 K, and of about 300 ps at $T$ = 1100 K.
The time interval between two uncorrelated rough configurations in MD
simulations is found to be of the same order.

\acknowledgements
We wish to thank Professor Robert Swendsen for useful suggestions.
This work has been partly supported by the Academy of Finland (JM)
and by the Emil Aaltonen Foundation (HH).
Part of the computations have been performed on Cray X-MP at the Center of
Scientific Computing (CSC) in Espoo, Finland.

\appendix
\section*{Cell method and neighbour list}
A well-known problem in MD simulations with short-range interactions is the
question of how to select only those pairs of atoms  which are within
the range of the potential.
The brute-force testing of all the ${1\over 2}N(N-1)$ pairs at each time step
is a prohibitive waste of time even in supercomputers,
if $N$ is of the order of few thousands atoms.
In our simulations,
we have constructed the so-called neighbour  list,\cite{Ho81}
which includes the indices $j$ of all neighbours of atom $i$ for which
$r_{ij} < r_L$, $r_L$ the list radius.
By using this list, only ${1\over 2}NN_L$ interactions have to be calculated,
where $N_L$ is the average number of neighbours for a given atom.
The cutoff $r_C$ in the EMT interaction potential\cite{Ha92} is between
the third and fourth nearest neighbours in the $T=0$ fcc copper lattice.
We have set $r_L$  to correspond to the distance to the middle of the
sixth and seventh neighbours,
which means that $N_L$ falls within the range of 80-90.
In order to minimize the frequency of the updating of the list,
we study at each time step the displacements of atoms from the previous
updating step.
The list is updated  whenever the maximum displacement exceeds half of
the list 'skin', $r_L-r_C$.

For the sample sizes used in this work, also the $N^2$ computations needed
for updatings become significant.
We have managed to speed up the updating using the following method.
We divide the sample into $n_x\times n_y$ subcells according to the number
of the  2D lattice sites,
i.e.,  30x30 or 50x50 for the samples studied in this work.
For each subcell $c_i$,
the indices  of neighbouring subcells $c_j$ are stored in a map
in the beginning of the simulation.
After attaching a given atom to its subcell,
it is sufficient to go through its own subcell and (in our case)  only 15
neighbouring subcells to construct its  neighbour list.
This method, being  linear with the system size,
clearly overrides the $N^2$ searches over all pairs, and brings the time
needed for updating comparable to that needed in the calculation of
the EMT interactions.
Our method can be regarded as a 2D modification of the so-called linked-cell
method,\cite{Ho81} with the important difference that the size of the subcell
is chosen to accommodate {\it only one lattice site} in two dimensions.
In the present method - remember that we also use
a stable integration algorithm which allows a large time step - we have
been able to perform MD simulations for 30000
atoms (20000 of which dynamical) on a Decstation 5000/200  (3.7 Mflops)
scalar machine at a rate of about 20 ps/day.

\begin{figure}
\caption{Energy of an atom in EMT lattice-gas (full curves).
The broken curve is the derivative of $E$ with respect to $C_1$, and
the dotted line is the energy in a simple pair potential model.}
\label{fig1}
\end{figure}

\begin{figure}
\caption{Mapping of the Cu(110) surface to the six-vertex model.
(a) Correspondence between the fcc(110) face and the six-vertex model
(a model configuration with a vacancy on an otherwise flat surface), and
(b) energies of the six vertices allowed by the solid-on-solid restriction.}
\label{fig2}
\end{figure}

\begin{figure}
\caption{Height-height correlation functions for (a) lattice-gas model
and (b) six-vertex model.}
\label{fig3}
\end{figure}

\begin{figure}
\caption{Coefficient $A$ as a function of temperature
(a) from the correlation functions of the lattice-gas model,
(b) from the correlation functions of the six-vertex model  and
(c) from the interfacial width of the lattice-gas model.}
\label{fig4}
\end{figure}

\begin{figure}
\caption{(a) Formation of the first 'adsorption layer' in the lattice-gas model
 and (b) inverse slope of the adsorption isotherms at the layering transition
point.}
\label{fig5}
\end{figure}

\begin{figure}
\caption{Snapshots of lattice-gas configurations for two temperatures.}
\label{fig6}
\end{figure}

\begin{figure}
\caption{The occupation of surface layers $a_2,...,c_2$ as a function
of simulation time in MD runs A and B for $T$ = 1150 K.
Run A was initiated from a rough lattice configuration and run B from an
undefected smooth surface.}
\label{fig7}
\end{figure}

\begin{figure}
\caption{Snapshots of two-dimensional clusters in layer $a_1$
in MD for $T =$ 1150 K.}
\label{fig8}
\end{figure}

\begin{figure}
\caption{The probability distribution p(z;t) showing the atomic
mobility perpendicular to the (110) surface.}
\label{fig9}
\end{figure}

\begin{figure}
\caption{The probability distribution p(x;t) showing the atomic
mobility in the $\lbrack 001\rbrack$ direction parallel to the (110) surface.}
\label{fig10}
\end{figure}

\begin{figure}
\caption{The natural logarithm of the residence time distribution
for the surface layers. Shown also are  the linear fits,
from which the mean residence times $\tau$ for each layer have been estimated.}
\label{fig11}
\end{figure}

\begin{figure}
\caption{Size distribution of two-dimensional clusters in
layer $a_1$ (a) as a function of time in the MD run C
for $T$ = 1100 K and (b) in the MD run B for $T$ = 1150 K.
The dotted line in (b) is the LGMC behaviour at the same temperature.}
\label{fig12}
\end{figure}

\end{document}